\documentclass[aps,prb,showpacs,amsmath,amssymb,twocolumn,superscriptaddress]{revtex4}
\pdfoutput=1

\usepackage{graphicx}
\usepackage{color}
\usepackage{amsmath}

\begin{document}

\title{Non-universal shot noise in quasiequilibrium spin valves}

\author{T.~T.~Heikkil\"a}
\affiliation{Low Temperature Laboratory, Aalto University, P.O. Box 15100, FI-00076 AALTO, Finland}

\author{K.~E.~Nagaev}
\affiliation{Kotelnikov Institute of Radioengineering and Electronics, Mokhovaya 11-7, Moscow, 125009 Russia}
\affiliation{Moscow Institute of Physics and Technology, Institutsky per. 9, Dolgoprudny, 141700 Russia}

\newcommand{\tmpnote}[1]%
   {\begingroup{\it (FIXME: #1)}\endgroup}
   \newcommand{\comment}[1]%
       {\marginpar{\tiny C: #1}}

\date{\today}

\begin{abstract}
We show that the breakdown of the Wiedemann-Franz law due to electron--electron scattering in diffusive spin valves may result in a strong suppression of the Fano factor that describes the ratio between shot noise and average current. In the parallel configuration of magnetizations, we find the universal value $\sqrt{3}/4$ in the absence of a normal-metal spacer layer, but including the spacer leads to a non-monotonous suppression of this value before reaching back to the universal value for large spacer lengths. On the other hand, in the case of an antiparallel configuration with a negligibly small spacer, the Fano factor is $\sqrt{3 (1-P^2)}/4$, where $P$ denotes the polarization of the conductivities. For $P\rightarrow \pm 1$, the current through the system is almost noiseless.
\end{abstract}
\pacs{75.60.Jk, 72.15.Jf, 75.30.Sg, 85.75.d}

\maketitle
\section{introduction}

Shot noise is often used as a tool for characterizing the magnitude of energy relaxation for example due to electron-phonon scattering in small electron systems (for a recent example, see Ref.~\onlinecite{Betz:2012dy}). This method relies on the fact that the noise power $S_I(\omega)$, defined as the Fourier transform of the current-current correlator, through a given conductor is quite generally proportional to the average of the electron temperature in this sample. Biasing the system with a voltage $V$ induces Joule heating, and the static electron temperature follows from a balance between this heating and the energy relaxation out of the system. This way the strength of energy relaxation can be read off from the noise.

At low temperatures in metals the main energy relaxation mechanism is due to the direct diffusion of electrons to the electrodes. This heat conduction is typically characterized by the Wiedemann-Franz (WF) law,\cite{Franz:1853ek} according to which the heat conductivity $\kappa={\cal L}_0 \sigma T$ is directly proportional to the charge conductivity $\sigma$ and the local electron temperature $T$. The proportionality factor ${\cal L}_0=\pi^2 k_B^2/(3e^2)$ is known as the Lorenz number. In the limit of a large voltage compared to the temperature $T_0$ of the electrodes, $|eV|\gg k_B T_0$, the WF law results into shot noise which is proportional to the average current $\langle I \rangle$, with a universal Fano factor $F\equiv S_I/(2e |\langle I \rangle|) = \sqrt{3}/4$.\cite{Nagaev:1995ju,Sukhorukov:1999bp} This Fano factor is independent of the sample geometry and applies even in the presence of a non-uniform electron conductivity or a non-uniform cross-section of the sample. The only requirements are the validity of the WF law and the quasiequilibrium limit,\cite{Wellstood:1994ec} where the electron--electron scattering is strong enough, so that the electron distribution function can everywhere be described via a Fermi-Dirac distribution function with a well-defined position-dependent electron temperature. This value has been demonstrated experimentally\cite{Steinbach:1996vq} more than a decade ago and it is now well-established. 

\begin{figure}[t]
\centering
\includegraphics[width=0.9\columnwidth]{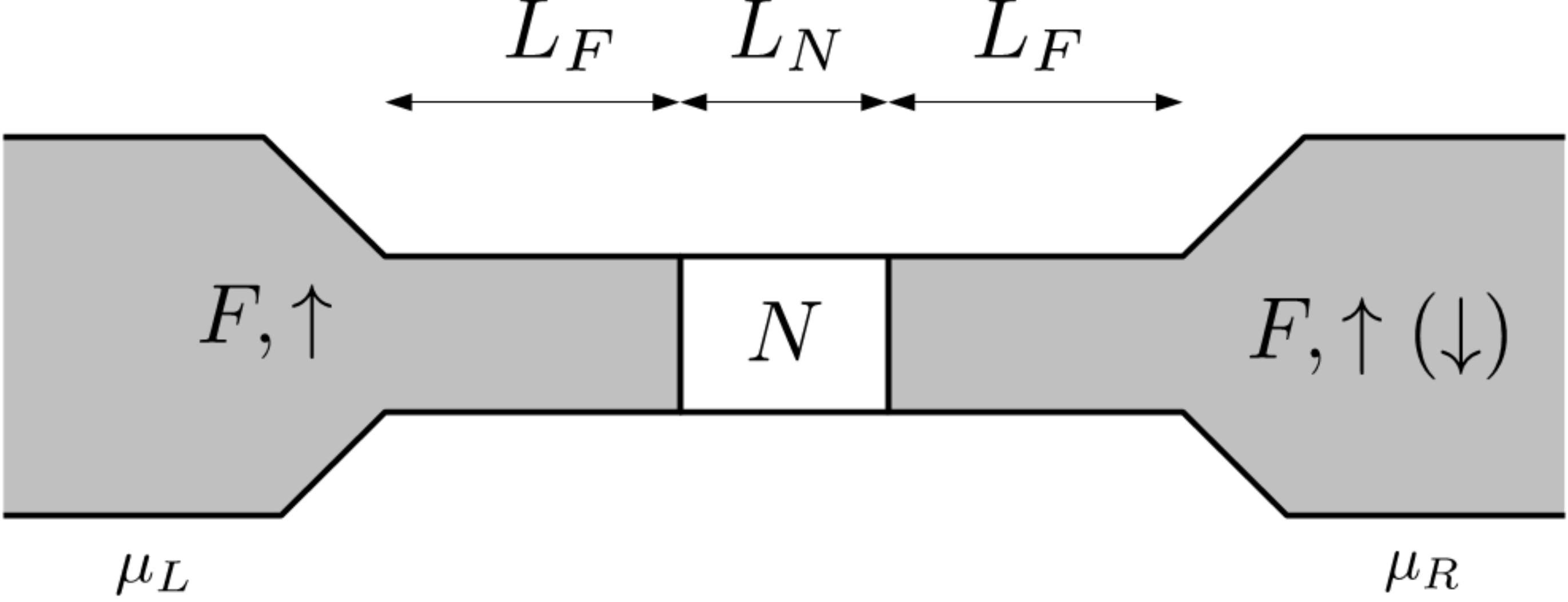}
\caption{Spin valve considered in this Letter: two ferromagnetic wires with length $L_F$ are connected to each other via a normal-metal spacer of length $L_N$ and to electrodes held at different potentials $\mu_L$ and $\mu_R=\mu_L-eV$ and at a temperature $T_0 \ll |eV|/k_B$.}
\label{fig:system}
\end{figure}

Recently a lot of interest has been devoted to the problem of heat transfer in magnetic systems.\cite{Bauer:2012fq} One of the key findings is that as internal energy relaxation of the electrons couples the energies in the two spin systems without relaxing their charges, the WF law breaks down.\cite{Hatami:2007gp,Heikkila:2010fg} 
In this paper we show how the breakdown of the WF law leads to a shot noise with a Fano factor deviating from the universal value in these devices. In particular, we find that in a spin valve with antiparallel orientation of magnetizations (Fig.~\ref{fig:system}), the Fano factor is strongly suppressed implying an almost noiseless transmission of current through the device. This is counterintuitive, as the antiparallel magnetization orientation can be envisaged as a reduced overall transmission of electrons through the spin valve --- compared to the parallel orientation --- and typically such a reduction of transmission leads to an increased Fano factor rather than a reduced one.\cite{Blanter:2000wi} This behavior also sharply contrasts with the results for spin valves with spin-flip scattering in the paramagnetic metal,\cite{Mishchenko03} where the maximum Fano factor corresponds to the maximum resistance. 


\begin{figure}[t]
\centering
\includegraphics[width=0.9\columnwidth]{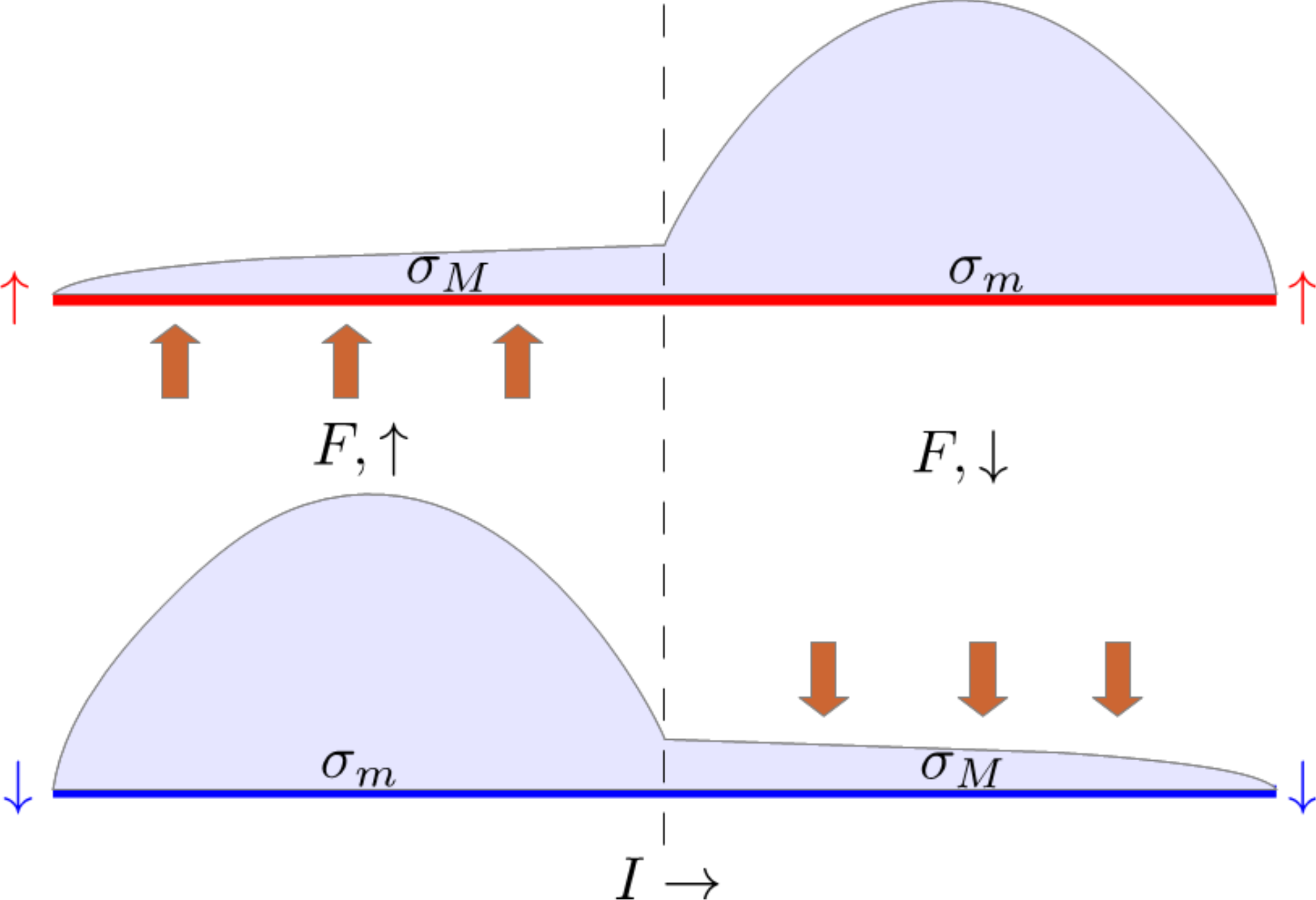}
\caption{(Color online): The would-be distribution of temperature along the spin valve for spin-up and spin-down electrons in the absence of interaction between them for the antiparallel magnetizations of the electrodes in the extreme case $L_N=0$. As the  higher temperature corresponds to the lower conductance, the would-be temperature profiles are highly asymmetric, and switching on the interaction results in a heat exchange whose direction is shown by orange arrows.}
\label{fig:isolated}
\end{figure}

If there were no spin-flip and electron-electron scattering, a diffusive spin valve could be considered as 
two isolated parallel conducting channels with spin-up and spin-down electrons connecting the same  reservoirs. Each channel 
would separately give a Fano factor $1/3$, yielding the same overall Fano factor characteristic of diffusive wires out of equilibrium. This value would be independent of electrode magnetizations or the details of the sample.\cite{Abdollahipour:2006jf,Tserkovnyak:2001jt}

If each spin subsystem were in local equilibrium but there were no heat exchange between electrons with opposite spins, this quantity would change to $\sqrt{3}/4$ and remain sample-independent because of the direct proportionality between the electric and heat conductivities.  However if the heat exchange between spin subsystems is switched on, it results in a heat transfer between the channels, which is not accompanied by any charge transfer and hence may violate the Wiedemann - Franz law. This increases the effective $\kappa/\sigma$ ratio and decreases the Fano factor below $\sqrt{3}/4$.

In the case of parallel magnetizations of the electrodes and a vanishing normal region, the spatial dependence of 
the electron temperature in both spin-up and spin-down channels would have the same symmetric shape and therefore would not be affected 
by heat exchange between the channels. However in the case of antiparallel magnetization, the distribution of temperature in thermally 
isolated channels would be strongly asymmetric (see Fig.\ref{fig:isolated}). The regions where the electrons with given spin direction are minority carriers 
would have a much lower conductivity and a much higher temperature than the ones where they are in the majority. Therefore it is the minority-carrier regions that would be the dominant sources of noise. As the heat exchange between the spin subsystems is switched on, the hotter minority-carrier regions appear to be in direct thermal contact with much colder majority carriers in the opposite-spin channel. As the majority carriers have much larger electric and thermal conductivity, they represent almost perfect heat sinks for the minority carriers  and strongly suppress their effective temperature together with the Fano factor.

\section{Model and basic equations}

We describe the ferromagnet--normal-metal--ferromagnet (FNF) system shown schematically in Fig.~\ref{fig:system}, where the ferromagnets of equal length $L_F$ are connected to a normal-metal spacer with length $L_N$ and conductivity $\sigma_N$. For simplicity we only consider the collinear orientation of magnetizations. In the case of parallel magnetizations, the majority (minority) spin ``up'' (``down'') electrons in both electrodes have conductivity $\sigma_M$ ($\sigma_m$) with $\sigma_m \le \sigma_M$, whereas in the case of antiparallel magnetization orientation the conductivities of the spin-up and spin-down channels are interchanged in the right ferromagnet. An alternative way to describe the spin-dependent conductivities is to define the average conductivity $\sigma_F \equiv (\sigma_M+\sigma_m)/2$ and spin polarization $P=(\sigma_M-\sigma_m)/(2\sigma_F)$. 

In the diffusive limit, where all length scales in the problem are large compared to the elastic mean free path $\ell_{\rm el}$, the electron distribution function $f_s(x,E)$ for spin $s \in \{\uparrow,\downarrow\}$ electrons satisfies \cite{Valet:1993es}
\begin{equation}
D_s \partial_x^2 f_s = I_{\rm e-e}^s[f_s,f_{\bar s}]+I_{\rm e-ph}^s[f_s,n_{\rm ph}],
\label{eq:boltzmann}
\end{equation}
where $D_s$ is the diffusion constant for spin $s$, $\bar s$ denotes the spin opposite to $s$, $I_{\rm e-e}$ and $I_{\rm e-ph}$ are the collision integrals for electron--electron (e-e) and electron--phonon scattering. In what follows, we concentrate on low temperatures where the latter can be disregarded.\cite{Giazotto:2006fm} The role of e-e scattering is to equilibrate the electron system into a common local temperature without any transfer of electron energy to non-electronic excitations. However, whereas in the absence of spin-flip scattering the spin-dependent potentials are unaffected by e-e scattering, the energies of the two spin ensembles are coupled\cite{Dimitrova:2008ed,Chtchelkatchev:2008km,Heikkila:2010fg,Heikkila:2010db} and therefore in the limit of strong e-e scattering spin-up and spin-down electrons can be described with the same temperature $T(x)$.

In the quasiequilibrium limit the distribution hence reads $f_s(x,E)=f_0(E;\mu_s(x),T(x))$, where $f_0(E;\mu,T)=\{\exp[(E-\mu)/(k_B T)]+1\}^{-1}$ is the Fermi function. The spin-dependent potentials satisfy the continuity equation
\begin{equation}
\partial_x [\sigma_s \partial_x \mu_s(x)]=0,
\label{eq:potentialequation}
\end{equation}
which is obtained by integrating Eq.~\eqref{eq:boltzmann} over the energy.
We assume a position-dependent conductivity $\sigma_s=e^2 D_s N_s$, where $N_s$ is the density of states for spin $s$. The vanishing energy flux out of the electron system due to e-e scattering requires that $\sum_s N_s \int dE E I_{\rm e-e}^s(E)=0$. 
With the help of this equality we obtain the heat diffusion equation for temperature $T(x)$ by multiplying Eq.~\eqref{eq:boltzmann} by $N_s$ times energy, summing over spin and integrating over the energy. This yields
\begin{equation}
{\cal L}_0 \partial_x \left[\left(\sum_s \sigma_s\right) T \partial_x T\right]
=-\frac{1}{e^2}\sum_s \sigma_s \left(\partial_x \mu_s\right)^2.
\label{eq:heatdiffusion}
\end{equation}
This equation describes the diffusion of the thermal energy (left hand side) in response to Joule heating (right hand side) generated throughout the sample for both spins.

As we assume the interface resistances to be negligible as compared with those of the wires, the boundary conditions for Eqs.~\eqref{eq:potentialequation} and \eqref{eq:heatdiffusion} consist in  continuity of the potentials and the temperature and in conservation of the heat current and spin-resolved charge currents (terms in square brackets on the left-hand-sides of Eqs.~\eqref{eq:potentialequation} and \eqref{eq:heatdiffusion}) across the interfaces.

Once the potential profile has been found, the position-independent average current is easily calculated as
\begin{equation}
\langle I \rangle = A \sum_s \sigma_s \int dE \partial_x f_s(x,E) = A \sum_s \sigma_s \partial_x \mu_s,
\label{eq:observables}
\end{equation}
where $A$ is the cross-section of the sample and the sum goes over the two spin directions. The Langevin equation 
for the current fluctuation\cite{Sukhorukov:1999bp} may be written in the form
\begin{equation}
 \delta I = \sum_s 
 \left[A\sigma_s\partial_x\delta\mu_s + \delta I_s^{ext}(x)\right],
 \label{eq:Langevin}
\end{equation}
where the correlation function of Langevin sources is given by
\begin{equation}
 \langle\delta I_s^{ext}(x)\,\delta I_{s'}^{ext}(x')\rangle
 =4A\sigma_s\,\delta_{ss'}\,\delta(x-x')\,\Lambda_s(x),
 \label{eq:sources}
\end{equation}
and $\Lambda_s(x) = \int dE f_s(x,E)(1-f_s(x,E))$. In the quasiequilibrium limit $\Lambda_s(x)=k_B T(x)$ is independent of spin. As $\delta I$ is independent of $x$ in the low-frequency limit\cite{Nagaev92}, the zero-frequency current noise power can be easily obtained from Eq.~(\ref{eq:Langevin}) in the form
\begin{equation}
S_I=4A\sum_s \frac{\int dx \Lambda_s(x)/\sigma_s(x)}{[\int dx \sigma_s^{-1}]^2}.
\label{eq:shotnoise}
\end{equation}

\begin{figure}[t]
\centering
\includegraphics[width=\columnwidth]{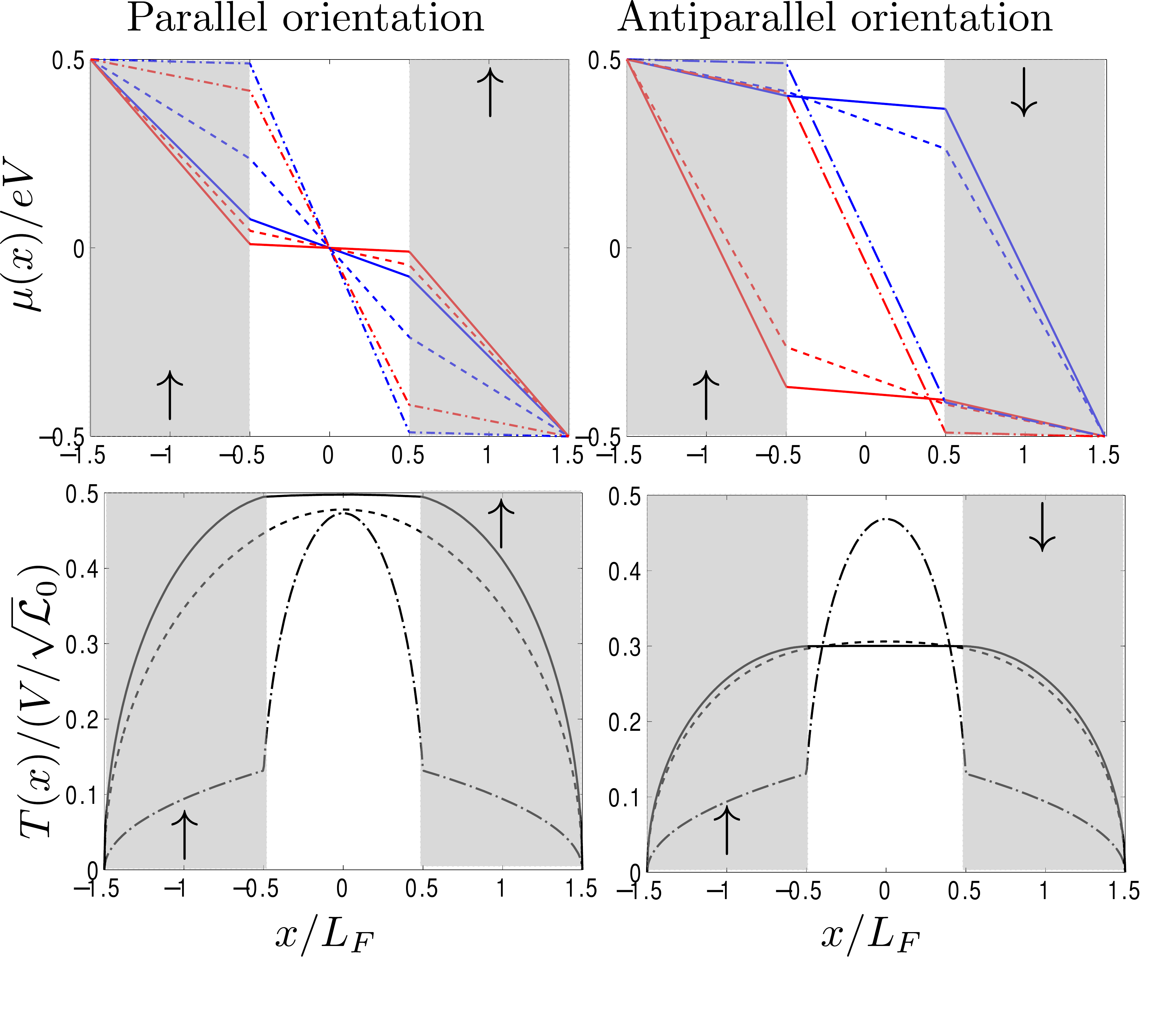}
\caption{(Color online): Potential (top) and temperature (bottom) profiles in spin valves with different magnetization configurations. The shaded regions refer to the ferromagnets, where the magnetization directions are denoted by the arrows. In the top figures, the blue curves (upper on the left part of the figure) are for $\mu_\uparrow$, whereas the red (lower on the left) are for $\mu_\downarrow$. The three sets of curves are for $\sigma_F/\sigma_N=0.2$ (solid lines), $\sigma_F=\sigma_N$ (dashed lines) and $\sigma_F=50 \sigma_N$ (dash-dotted lines). We have chosen $L_N=L_F$ and $P=0.8$.}
\label{fig:potentialandtemperatureprofile}
\end{figure}

The solution of the quasiequilibrium equations, Eqs.~\eqref{eq:potentialequation} and \eqref{eq:heatdiffusion} is straigthforward but somewhat lengthy. 
The obtained potential and temperature profiles are plotted for a few example cases in Fig.~\ref{fig:potentialandtemperatureprofile}.

\begin{figure}[t]
\centering
\includegraphics[width=0.8\columnwidth]{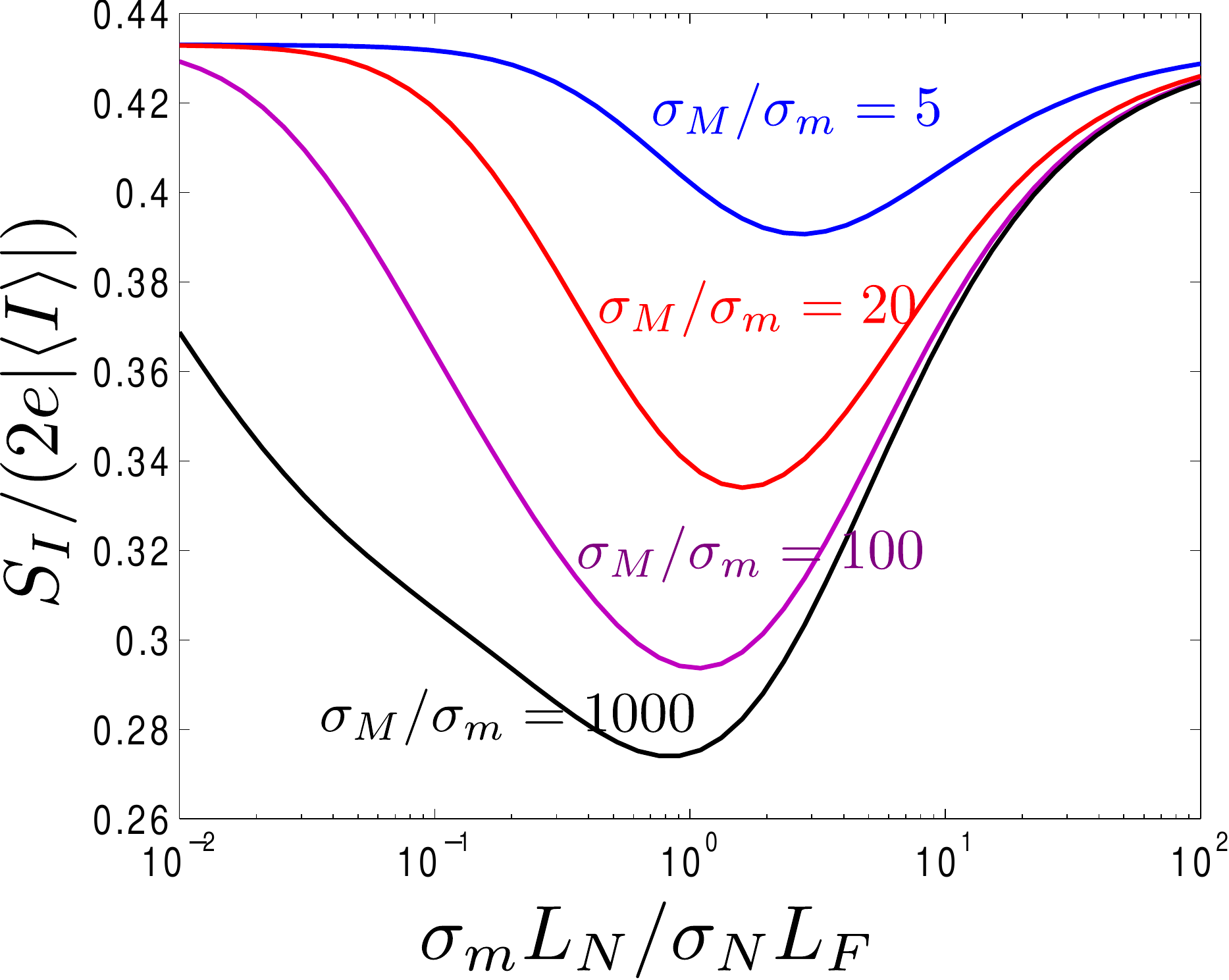}
\caption{(Color online): Fano factor of the spin valve for parallel magnetizations as a function of the ratio of minority electron conductance $\sigma_m A/L_F$ to the normal-metal conductance $\sigma_N A/L_N$ for different ratios $\sigma_M/\sigma_m$. Note the log-scale in the horizontal axis. For $\sigma_N A/L_N \rightarrow 0$ the normal metal dominates the resistance, and the Fano factor tends to the value $\sqrt{3}/4\approx 0.433$. In the opposite limit $\sigma_N/L_N \gg \sigma_M/L_F$ a symmetric spin valve does not maintain any spin accumulation, and the same Fano factor $\sqrt{3}/4$ is retained. }
\label{fig:parallelfano}
\end{figure}

With the knowledge of the temperature, the shot noise is obtained from Eq.~(\ref{eq:shotnoise}).
The expressions for the Fano factor for an arbitrary range of parameters are too lengthy to be presented here, but in the following we consider some simpler limits and present the more general results in Figs.~\ref{fig:parallelfano} and \ref{fig:antiparallelfano}. In general, the Fano factor depends on the magnetization configurations (parallel or antiparallel), spin polarization $P$, and the ratio of the spin-averaged conductances of the ferromagnets and the normal-metal spacer, i.e., $\alpha \equiv \sigma_F L_N/(\sigma_N L_F)$. Alternatively, we can describe this dependence with the parameters $\alpha_{M/m}=\sigma_{M/m} L_N/(\sigma_N L_F)=(1\pm P)\alpha$.

\section{Parallel orientation}

In the absence of spin heat accumulation, the validity of the Wiedemann-Franz law depends on the presence or absence of spin accumulation. The latter vanishes if $\sigma_M=\sigma_m$, i.e., $P=0$, and both in the limit where majority of the resistance comes from the normal-metal spacer, i.e., $\alpha \gg 1$ or in the opposite limit $\alpha \ll 1$. In the latter case the absence of spin accumulation results from the symmetry of the setup. The temperature distributions in the thermally uncoupled spin-up and spin-down channels would be identical and therefore would not be affected by the heat exchange.
In those limits we hence obtain the universal result $F_P=\sqrt{3}/4$ for the Fano factor. The intermediate case is plotted in Fig.~\ref{fig:parallelfano}. We find that $F$ obtains a minimum when $\alpha_m \approx 1$ and this minimum becomes wider and deeper as the polarization increases. For $\alpha_M \rightarrow \infty$ the curve approaches its limiting shape given by
\begin{equation}
F_P \overset{\alpha_M \gg \alpha_m,1}{\longrightarrow} \frac{\sqrt{3}}{4} \frac{(\alpha_m^2+2\alpha_m +2)^{3/2}}{(\alpha_m+1)(\alpha_m+2)^2}.
\label{eq:parallelfano}
\end{equation}
Hence $F \approx \sqrt{6}/8$ in a wide range $\alpha_m \ll 1 \ll \alpha_M$. The minimal value $F = 9\sqrt{71-17^{3/2}}/32 \approx 0.27$ is obtained at $\alpha_m = (\sqrt{17}-1)/4$. Note that this limit requires quite a strong polarization of the ferromagnets (see Fig.~\ref{fig:parallelfano}). 

\section{Antiparallel orientation}

In the antiparallel configuration of the magnetizations, the spin accumulation is non-zero even in the case of a negligible resistance of the normal-metal spacer (i.e., $\alpha \ll 1$). In this case 
\begin{equation}
F_{AP} \overset{\alpha \ll 1}{\rightarrow} \frac{\sqrt{3 \sigma_M \sigma_m}}{2(\sigma_M+\sigma_m)}=\frac{\sqrt{3}}{4} \sqrt{1-P^2}.
\label{eq:F_AP}
\end{equation}
The shot noise power is therefore suppressed by inter-spin relaxation due to e-e scattering. The quasiequilibrium Fano factor tends below the nonequilibrium limit $F_{AP}=1/3$ when $P>\sqrt{11/27} \approx 0.64$ and the shot noise (almost) vanishes in the half-metal limit $P \rightarrow 1$ (note that also the average current vanishes there, but the noise vanishes faster). The reason is that the  distribution of temperature in the case of a thermally uncoupled spin-up and spin-down channels would be strongly asymmetric in this case. The electron-electron scattering balances these temperatures, resulting in a strongly suppressed average temperature together with a suppressed Fano factor.

\begin{figure}[t]
\centering
\includegraphics[width=0.8\columnwidth]{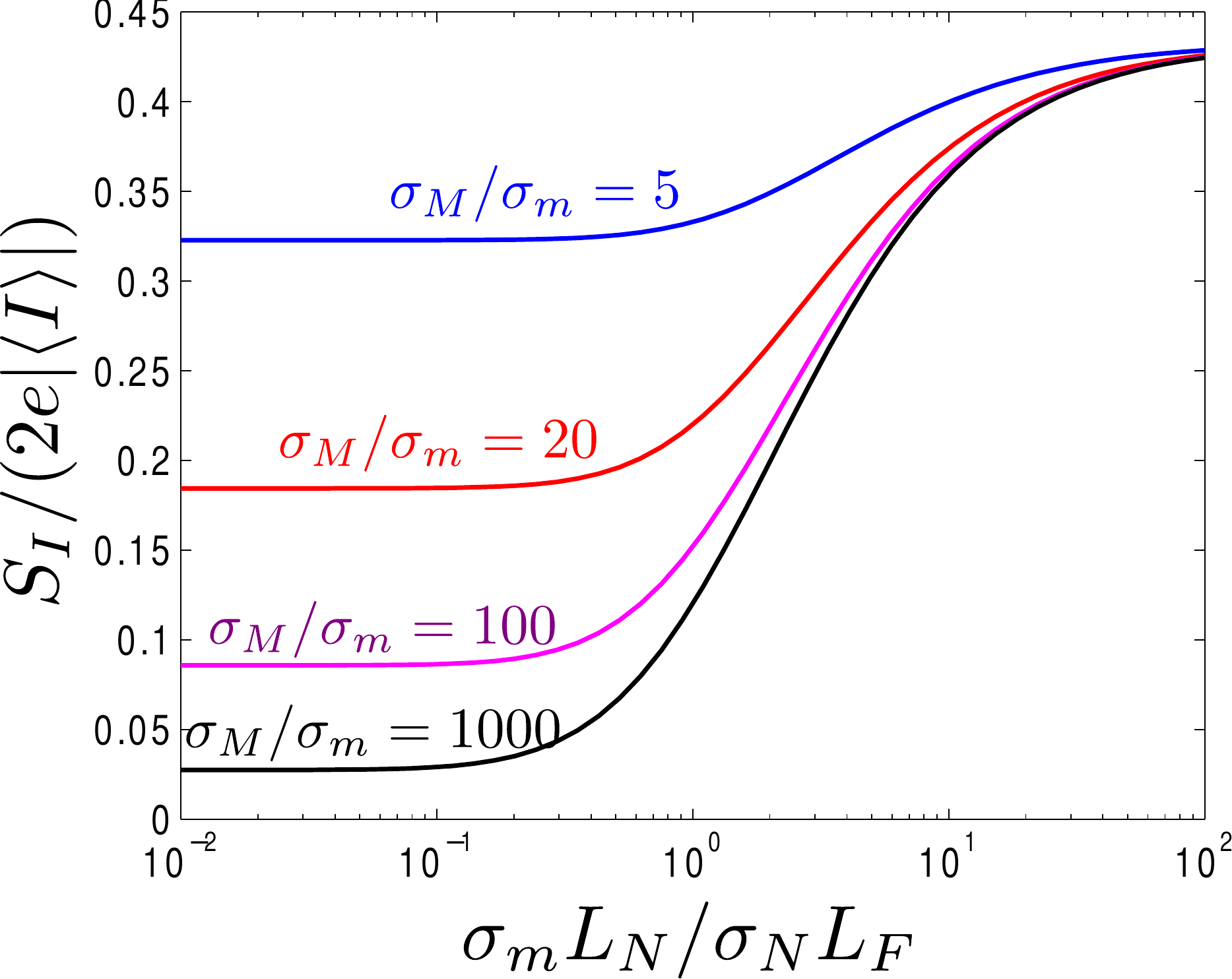}
\caption{(Color online): Fano factor for antiparallel magnetizations as a function of the ratio of the conductance for minority electrons to the conductance of the normal metal, for different ratios $\sigma_M/\sigma_m$. Note the log-scale in the horizontal axis. For $\sigma_N A/L_N \to 0$ the normal metal dominates the resistance, and the Fano factor tends to  $\sqrt{3}/4$. In the opposite limit $\sigma_N/L_N \gg \sigma_M/L_F$ the Fano factor tends to $F_{AP}\le\sqrt{3}/4$ given by Eq.~(\ref{eq:F_AP}). }
\label{fig:antiparallelfano}
\end{figure}

As shown in Fig.~\ref{fig:antiparallelfano}, the normal-metal spacer limits the suppression of shot noise. For $\alpha_M \gg 1$ (or $P \rightarrow 1$) we obtain
\begin{equation}
F_{AP} \overset{\alpha_M \gg 1}{\longrightarrow} \frac{\sqrt{3}}{4} \frac{\alpha_m^2}{(1+\alpha_m)^2}.
\label{eq:antiparallelfano}
\end{equation}
This function thus interpolates between the full suppression of noise at $\alpha_m \ll 1$ and the normal quasiequilibrium limit $F=\sqrt{3}/4$ for $\alpha_m \gg 1$. 

\section{Discussion}

Let us now estimate the parameter regime where our discussion is valid. We have assumed that the length $L=2L_F+L_N$ is shorter than the spin diffusion length. Spin-flip scattering would suppress the spin accumulation, and therefore also the violation of the Wiedemann-Franz law. The deviations of the Fano factor from $\sqrt{3}/4$ would therefore be smaller in systems with stronger spin-flip scattering. On the other hand, the quasiequilibrium limit generally requires $\ell_{ee} \ll L_F, L_N \ll \ell_{e-ph}$, where $\ell_{e-e}$ is the electron-electron scattering length and $\ell_{e-ph}$ is the electron-phonon scattering length.  We may now generalize the overall behavior of the Fano factor from Ref.~\onlinecite{Steinbach:1996vq} to spin valves: for $L \ll \ell_{e-e}$, the Fano factor is 1/3, independent of the magnetization configuration or the microscopic details. Increasing the voltage and thereby decreasing $\ell_{e-e}$, Fano factor tends to another value, which now depends on both the magnetization configuration and the relative sizes of the spacer and the ferromagnets. 

\section{Summary}

In summary, we showed that the breakdown of the Wiedemann-Franz law via inter-spin energy relaxation leads to a drastic change 
in the shot noise through a diffusive spin valve. Unlike spin-flip scattering, the heat exchange between the two spin subsystems leads to
a suppression of the noise for the antiparallel magnetization of the electrodes when the resistance of the valve is maximal. This effect
may be used for determining the parameters of electron-electron scattering in ferromagnets.

\begin{acknowledgments}
This work was supported by the Academy of Finland, the European Research Council (Grant No. 240362-Heattronics) and EU-FP 7 MICROKELVIN program (Grant No. 228464). We thank P. Virtanen for fruitful discussions.
\end{acknowledgments}

\begin{widetext}

\appendix

\section{Full Fano factors}

For reference, we write here the full analytic forms for the Fano factors. In the parallel case the Fano factor is of the form
\begin{equation}
F_P=\frac{\sqrt{3}}{4\pi(\alpha_m+\alpha_M+\alpha_m\alpha_M)}
\left[A_1+\frac{2A_3 \sqrt{A_2}}{(2+\alpha_m)^2(2+\alpha_M)^2(\alpha_m+\alpha_M)}\arcsin\left(\sqrt{\frac{(\alpha_m+\alpha_M)A_2}{A_3}}\right)\right],
\end{equation}
where $A_1$, $A_2$, and $A_3$ are given by equations
\begin{align*}
A_1=&\frac{8 \left(\alpha _m+\alpha _m \alpha _M+\alpha _M\right){}^2} 
   {\sqrt{\left(\alpha _m+\alpha _M\right) 
          \left[\alpha _m \alpha _M \left(\alpha _m+\alpha _M+8\right)+4(\alpha_m+ \alpha_M)\right]}}
   \arcsin\!
   \left( \sqrt{\frac{\alpha _m \alpha _M \left(\alpha _m+\alpha _M+8\right)+4(\alpha_m+ \alpha_M)}
               {2\left(\alpha _m+2\right) \left(\alpha_M+2\right) \left(\alpha _m+\alpha _m \alpha _M+\alpha _M\right)}
          }
   \right),\nonumber\\
A_2=&2\alpha_M^2 + 2 \alpha _m^2 + \alpha_m \alpha_M (2 \alpha_M+\alpha_m \alpha_M + 2 \alpha_m),\\
A_3=&
   \alpha _m^3 \left(\alpha _M \left(\alpha _M+2\right)+2\right)+\alpha _m^2 \left(\alpha _M \left(\alpha _M \left(\alpha
   _M+12\right)+22\right)+16\right)+2 \alpha _m \left(\alpha _M \left(\alpha _M \left(\alpha _M+11\right)+16\right)+8\right)\\&+2 \alpha _M
   \left(\alpha _M \left(\alpha _M+8\right)+8\right)\nonumber
\end{align*}
%
In the antiparallel case, the Fano factor is of the form
%
\begin{equation}
F_{\rm AP} = \frac{\sqrt{3}\,(N_1+N_2)}{D},
\end{equation}
%
where $N_1$, $N_2$, and $D$ are given by equations
%
\begin{align*}
N_1&=4\,\sqrt{\alpha_m\alpha_M}\,
   (\alpha_m+\alpha_M)(\alpha_m+\alpha_M+\alpha_m\alpha_M)^2
   \arcsin\!\left(\sqrt{\frac{\alpha _m+\alpha _M}
                           {2(\alpha_m+\alpha _m \alpha _M+\alpha _M)}}
          \right),
          \\
N_2&=\alpha _m \alpha _M
   \left(\alpha _m+\alpha _M\right) 
   \left[\alpha _m \alpha _M \left(\alpha _m+\alpha _M+8\right)+4\,(\alpha_m+\alpha_M)\right] 
   {\rm arccot}\!
   \left(2\sqrt{\frac{(\alpha _m+2 \alpha _m \alpha _M+\alpha _M)}
                    {\alpha _m \alpha _M \left(\alpha _m+\alpha_M\right)}}
   \right),\\
D&=2  \pi  \left(\alpha _m+\alpha _M\right){}^2 \left(\alpha _m+\alpha _m \alpha
   _M+\alpha _M\right){}^2.
\end{align*}

\end{widetext}

\end{document}